\documentclass[showpacs,aps]{revtex4}
\usepackage{graphicx}
\usepackage{amsfonts}
\usepackage{multirow}
\usepackage{braket}
\usepackage[usenames,dvipsnames]{color}
 \usepackage{amsmath,amssymb,bm,euscript}
 \usepackage{graphicx,boxedminipage}
 \usepackage{floatflt}
 \usepackage[hypertex]{hyperref}
 \usepackage{xcolor,fancybox}

\definecolor{Dark-blue}{RGB}{0,0,255}
\newcommand{\av}[1]{\ensuremath{\langle#1\rangle}}
\newcommand{\Av}[1]{\ensuremath{\big<#1\big>}}

\reversemarginpar \marginparwidth=1cm

\begin{document}

\title{The plasmon-polariton mirroring due to strong fluctuations of the surface impedance}

\author{Yu.\,V.~Tarasov,  D.\,A.~Iakushev, S.\,S.~Melnik, O.\,V.~Usatenko}

\begin{abstract}
Scattering of TM-polarized surface plasmon-polariton waves (PPW) by
a finite segment of the metal-vacuum interface with randomly
fluctuating surface impedance is examined. Solution of the integral
equation relating the scattered field with the field of the incident
PPW, valid for arbitrary scattering intensity and arbitrary
dissipative characteristics of the conductive medium, is analyzed.
As a measure of the PPW scattering, the Hilbert norm of the integral
scattering operator is used. The strength of the scattering is shown
to be determined not only by the parameters of the fluctuating
impedance (dispersion, correlation radius and the length of the
inhomogeneity region) but also by the conductivity of the metal. If
the scattering operator norm is small, the PPW is mainly scattered
into the vacuum, thus losing its energy through the excitation of
quasi-isotropic bulk Norton-type waves above the conducting surface.
The intensity of the scattered field is expressed in terms of the
random impedance pair correlation function, whose dependence on the
incident and scattered wavenumbers shows that in the case of
random-impedance-induced scattering of PPW it is possible to observe
the effect analogous to Wood's anomalies of wave scattering on
periodic gratings. Under strong scattering, when the scattering
operator norm becomes large compared to unity, the radiation into
free space is strongly suppressed, and, in the limit, the incoming
PPW is almost perfectly back-reflected from the inhomogeneous part
of the interface. This suggests that within the model of a
dissipation-free conducting medium, the surface polariton is
unstable against arbitrary small fluctuations of the medium
polarizability. Transition from quasi-isotropic weak scattering to
nealy back-reflection under strong fluctuations of the impedance is
interpreted in terms of Anderson localization.
\end{abstract}

\maketitle

\section{Introduction}

The complexity of electromagnetic phenomena in the study of surface polariton waves has led in due time to the need for determining some simple parameters that would allow for easy modeling, calculations and design of electromag\-netic (EM) devices. Among these simple and reliable parameters the impedance is most commonly used. It allows to greatly simplify the solution of the problem and circumvents the necessity to study the solution of electrodynamic equations directly in a metal. A brief review of the work on the development of the impedance concept, different from that of Leontovich, can be found in paper~\cite{Kaiser2013}.

Regardless of the method of solving the problem --- either accurately taking into account the surface topography or modeling it through an impedance condition --- there appears a need to study an integral equation describing the PPW scattering processes. In particular, the integral equation arises from the use of the Leontovich boundary condition (functional equation) after substituting the solution of the Helmholtz equation in the form of an~expansion into eigen-waves (in the simplest case, plane waves) and switching to the momentum representation, see Eq.~\eqref{R(q)-mainEq} below or the paper by Depine~\cite{Depine92}.

\begin{figure}[h]
  \centering
  \scalebox{.7}[.7]{\includegraphics{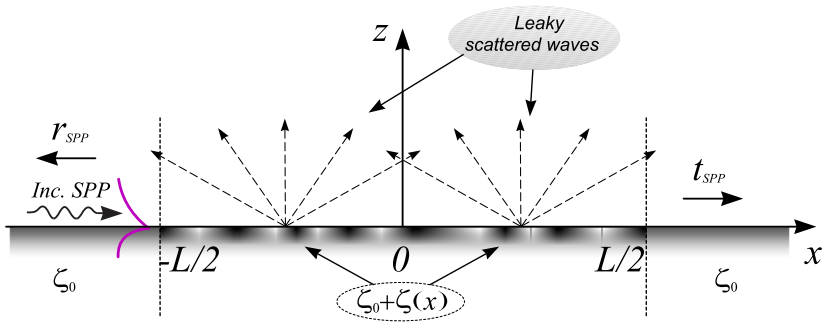}}
  \caption{The geometry of the PPW scattering on a flat boundary with the surface impedance randomly modulated over a segment of finite length $L$.
  \hfill\label{fig1}}
\end{figure}

In the present study, we examine plasmon-polariton scattering by a
metal strip of a finite width $L$ with fluctuating impedance (see
Fig.~\ref{fig1}). The wave vector of the plasmon-polariton lies in
the plane of the metal and is directed perpendicular to the
scattering section of the surface. The radiation arising due to the
fluctuations of the surface impedance is considered with the account
of the absorption due to the finite conductivity of the metal. For
the analysis of the problem, we apply a standard approach consisting
in the analysis of the integral equation for the scattered field
expressed through the Fourier components of the EM field on the
metal surface. The main difficulty in solving this equation is
associated with the presence of the integral term, which is usually
considered as a small perturbation, in powers of which the solution
is sought. We propose and justify the criterion for the evaluation
of the PPW scattering intensity, which is the Hilbert norm of the
mixing operator for the intermediate scattering states. A rough
qualitative assessment of the norm is that it is proportional to the
mean square fluctuation of the impedance and inversely proportional
to its real part (the accurate estimate is given by expression
\eqref{L_norm-def-2}). We calculated the scattered radiation pattern
and showed that for the weak scattering the radiated energy is
proportional to the Fourier transform of the binary correlation
function of the impedance, the argument of which is equal to the
difference between the wave number of the surface polariton and the
projection of the wave vector of the scattered field harmonic
directed towards the observation point. This dependence has an
obvious resemblance to the angular dependence of the field scattered
by reflective gratings, where Wood found the anomalies
\cite{Wood1902}.

We pay special attention to another limiting case, where the integral term in the master equation is major compared to those outside the integral. We show that with increasing the metal conductance, the fraction of the field scattered into vacuum increases compared to the PPW ohmic losses in the metal. Simultaneously, the contribution of the integral term grows with respect to that of the non-integral term. For strong impedance fluctuations, the scattered field becomes almost deterministic in nature, and in the limit when the dissipation of charge carriers in the metal completely vanishes, the PPW is totally reflected backwards by the surface region containing the fluctuating impedance. Thus, we show that in the model of perfectly conductive metal the integral term is the primary one, and the starting point for solving any equation containing inhomogeneity should be the solution of the integral rather than algebraic equation. In other words, in the model of a dissipation-free metal, the standard plasmon-polariton solution is unstable with respect to infinitesimal perturbations.

\section{Problem statement}
\label{Formulation}

Let us consider the two-dimensional problem of scattering of a surface plasmon-polariton wave excited by a certain
source (e.g., a slit~\cite{bib:Tejeira05,bib:Tejeira07}) on the impedance boundary between a metal and vacuum. A finite section of the boundary is assumed to have random impedance. We assume also that the impedance depends on one coordinate only and describe it with a function consisting of two complex terms,
\begin{equation}\label{Z}
  Z_s(x)=\zeta_0+\zeta(x)\ ,
\end{equation}
of which the first term, $\zeta_0$, is constant while the second term, $\zeta(x)$, is a random function of the $x$-coordinate. This function is assumed to have
non-zero value solely in the interval $\mathbb{L}:\,x\in[-L/2,L/2]$ and to have zero average value within this interval, $\av{\zeta(x)}=0$.

The surface plasmon-polariton is a $p$-polarized (TM) wave with the only nonzero component of the magnetic field $H_y\equiv H(\mathbf{r})$ that satisfies Helmholtz equation
\begin{subequations}\label{initial_eq_Psi}
\begin{equation}\label{Poisson_eq}
   \left(\Delta+k^2\right)H(\mathbf{r})=0\ ,
\end{equation}
and boundary condition
\begin{equation}\label{Impedance_BC}
  \left(\frac{\partial H}{\partial z}+ik\big[\zeta_0+\zeta(x)\big]H\right)
  \Bigg|_{z=0}=0\ .
\end{equation}
\end{subequations}
Symbol $\Delta$ in \eqref{Poisson_eq} denotes the (two-dimensional) Laplace operator, ${k=\omega/c}$ and $\mathbf{r}=(x,z)$ are the wavenumber and two-dimensional radius-vector, respectively. Due to the complexity of random function $\zeta(x)$ we set its binary correlation properties by Eqs.~\eqref{Basic_corrs(x)} below.

For the coordinate-independent surface impedance, problem \eqref{initial_eq_Psi} is easily solved and provided $\mathrm{Im}\,\zeta_0<0$ has, in general, two solutions of the following form,
\begin{align}\label{SPP}
  H_{\text{\tiny SPP}}^{(\pm)}(\mathbf{r})=& \mathcal{A}^{(\pm)}
  \exp\left(\pm ik_{\text{\tiny SPP}}x-i\zeta_0 kz\right)\ ,\\
\notag
  k_{\text{\tiny SPP}}=& k\sqrt{1-\zeta_0^2}\ .
\end{align}
We are interested in polaritons propagating \emph{quasi-freely} along the interface between two media for distances much greater than their wavelength. As it follows from Eq.~\eqref{SPP}, for this the following inequality should hold,
\begin{equation}\label{freeSPP}
  \zeta'_0\ll |\zeta''_0|\ .
\end{equation}
For real metals, the imaginary part of the impedance is negative, and its absolute value does not exceed unity (see, e.g., Ref.~\cite{Palik98}). The attenuation length of PPW due to the dissipation in the metal halfspace under condition \eqref{freeSPP} is represented by the following formula,
\begin{equation}\label{L(SPP)_dis}
  L^{\scriptscriptstyle{(\text{SPP})}}_{dis}= \big|k''_{\text{\tiny SPP}}\big|^{-1}
  = \frac{\sqrt{1+{\zeta_0''}^2}}{k\zeta_0'|\zeta_0''|}\ .
\end{equation}

Assuming that all the conditions necessary for the existence of a quasi-free PPW in the case of a homogeneous impedance ($\zeta(x)\equiv 0$) are fulfilled, we seek a solution of the problem \eqref{initial_eq_Psi} as a sum of certain unperturbed plasmon-polariton-type solution $H_0(\mathbf{r})$ and ``scattered'' field $h(\mathbf{r})$,
\begin{align}\label{h-BC}
  H(\mathbf{r})=H_0(\mathbf{r})+h(\mathbf{r})\ .
\end{align}
It is convenient to model the solution $H_0(\mathbf{r})$ in the form commonly adopted for one-dimensional scattering problems. In the region $x<-L/2$ we take it in the form of a sum of a plasmon-polariton incident from the left, which has the unit amplitude at the left point of the segment $\mathbb{L}$, and the reflected PPW, which propagates and is dissipatively attenuated in the opposite direction,
\begin{subequations}\label{H_0^lrc}
\begin{align}\label{H_0^<}
  H_0^{(l)}(\mathbf{r})=\exp\left[ik_{\text{\tiny SPP}}\big(x+ L/2\big)-i\zeta_0 kz\right]+
  r_-\exp\left[-ik_{\text{\tiny SPP}}\big(x+ L/2\big)-i\zeta_0 kz\right]\ .
\end{align}
To the right of $\mathbb{L}$, we choose the unperturbed field $H_0(\mathbf{r})$ in the form of transmitted PPW only,
\begin{align}\label{H_0^>}
  H_0^{(r)}(\mathbf{r})=t_+\exp\left[ik_{\text{\tiny SPP}}\big(x- L/2\big)-i\zeta_0 kz\right]\ .
\end{align}
And finally, in the intermediate region $x\in\mathbb{L}$, we set the unperturbed field in the form of a superposition of the field transmitted through the left boundary of $\mathbb{L}$ and the PPW reflected from its right boundary,
\begin{align}\label{H_0^in}
   H_0^{(in)}(\mathbf{r})=& t_-\exp\left[ik_{\text{\tiny SPP}}\big(x+ L/2\big)-i\zeta_0 kz\right]\nonumber\\
   &+ r_+\exp\left[-ik_{\text{\tiny SPP}}\big(x- L/2\big)-i\zeta_0 kz\right]\ .
\end{align}
\end{subequations}
Note here an excessive amount of arbitrary constants in the ``seed''
solutions~\eqref{H_0^lrc}. Below we impose two additional conditions
(in the form of continuity of the $y$-components of the scattered
magnetic field), which eliminate this redundancy.

The scattered field $ h(\mathbf{r})$ from representation
\eqref{h-BC} satisfies the Helmholtz equation \eqref{Poisson_eq},
but with the boundary condition on the surface $z=0$, which is
different from the condition \eqref{Impedance_BC}, viz.,
\begin{align}\label{psi(x)-BC}
  \left(\frac{\partial h(\mathbf{r})}{\partial z}+ik\big[\zeta_0+\zeta(x)\big]h(\mathbf{r})\right)
   \Bigg|_{z=0}= & -ik\zeta(x)
   \Big\{t_-\exp\left[ik_{\text{\tiny SPP}}\big(x+L/2\big)\right]\nonumber\\
   &+r_+\exp\left[-ik_{\text{\tiny SPP}}\big(x- L/2\big)\right] \Big\}\ .
\end{align}
Moreover, since in the regions $|x|>L/2$ condition \eqref{psi(x)-BC} allows for the wave equation solution in the form of surface polaritons only, and such solutions have already been taken into account in Eqs.~\eqref{H_0^lrc}, we can assume that in these areas the function $h(\mathbf{r})$ is identically equal to zero at $z=0$. As for the conditions which the scattered field should satisfy in the region $z>0$, given the finite size of the surface section considered as the source of this field, we assume that at large distances the field $h(\mathbf{r})$ fulfills the radiation conditions.

The choice of the unperturbed solution for the Helmholtz equation in the form of the set of functions \eqref{H_0^lrc}
associated with different sections of the $x$-axis allows us to couple the elements of the scattering matrix of a plasmon-polariton with its radiation in vacuum and dissipation in the conductive substrate. The dissipative losses are incorporated into the imaginary part of the surface plasmon wavenumber, $k_{\text{\tiny SPP}}$. As for the radiation losses, they are determined by the harmonics of the field $h(\mathbf{r})$ which leak into the upper half-space. If the field $h(\mathbf{r})$ is found, then by joining the values of the total field at the metal-vacuum interface at the inner side of the end points of the segment  $\mathbb{L}$ with the plasmon-polariton solutions \eqref{H_0^lrc} taken along the same border at points ${x=\pm (L/2+0)}$, we obtain the following set of equations relating the scattering coefficients $r_{\pm}$ and $t_{\pm}$,
\begin{subequations}\label{Joining_Eqs}
\begin{align}
\label{Join_relations-1}
 1+r_- &= t_- + r_+\mathrm{e}^{ik_{\text{\tiny SPP}}L} + h(-L/2,0)\ ,\\
\label{Join_relations-2}
 t_+ &= t_-\mathrm{e}^{ik_{\text{\tiny SPP}}L}+r_+ + h(L/2,0)\ .
\end{align}
\end{subequations}
From these equations it follows that the scattering coefficients of the ``unper\-turbed'' PPW appearing in Eqs.~\eqref{H_0^lrc} are directly related to the yet unknown scattered field $h(\mathbf{r})$, so they should be determined self-consistently.

\section{General solution for the scattered field}
\label{Full_Sol}

To find the field $h(\mathbf{r})$, we do not restrict ourselves to specific regions on the $x$-axis and consider the perturbed section of the conductor as a source of the field for the entire upper half-space. Let us seek for the scattered field in the form of an integral
\begin{equation}\label{Full_solution_h(r)}
  h(\mathbf{r})=\int\limits_{-\infty}^{\infty}\frac{dq}{2\pi}
  \widetilde{\mathcal{R}}(q)\exp\left[iq x+i\big(k^2-q^2\big)^{1/2}z\right]\ ,
\end{equation}
which satisfies by construction both the wave equation in free space and the radiation conditions at infinity. To determine the scattering amplitude $\widetilde{\mathcal{R}}(q)$, we use boundary condition \eqref{psi(x)-BC} making its Fourier transform with respect to~$x$. In this way we arrive at the following integral equation,
\begin{align}\label{R(q)-mainEq}
  \left[\zeta_0+\sqrt{1-(q/k)^2}\right] & \widetilde{\mathcal{R}}(q)
  +\int\limits_{-\infty}^{\infty}\frac{dq'}{2\pi}
  \widetilde{\mathcal{R}}(q')\widetilde{\zeta}(q-q') =\nonumber\\
 & -\mathrm{e}^{ik_{\text{\tiny SPP}}L/2}\left[t_-\widetilde{\zeta}\big(q-k_{\text{\tiny SPP}}\big)+
  r_+\widetilde{\zeta}\big(q+k_{\text{\tiny SPP}}\big)\right]\ .
\end{align}

If the real part of the surface impedance is not equal to zero ($\zeta_0'\neq 0$), the expression in the square brackets on the left-hand side of Eq.~\eqref{R(q)-mainEq} does not vanish on the real frequency axis. By dividing both sides of the equation by this factor, we can rewrite it in the form of the standard Fredholm integral equation of the second kind \cite{bib:Courant_Hilbert66},
\begin{subequations}\label{Fredholm}
\begin{align}\label{Fredholm_eq}
  \widetilde{\mathcal{R}}(q)+
  \int\limits_{-\infty}^{\infty}\frac{dq'}{2\pi}\mathcal{L}(q,q')\widetilde{\mathcal{R}}(q')
  =-\mathrm{e}^{ik_{\text{\tiny SPP}}L/2}\big[t_-\mathcal{L}(q,k_{\text{\tiny SPP}})+
  r_+\mathcal{L}(q,-k_{\text{\tiny SPP}})\big]\ .
\end{align}
The kernel of the integral operator on the left-hand side of Eq.~\eqref{Fredholm_eq} has the form
\begin{equation}\label{Kernel_Fredholm_eq}
  \mathcal{L}(q,q')=\left[\zeta_0+\sqrt{1-(q/k)^2}\right]^{-1}\widetilde{\zeta}(q-q')\ .
\end{equation}
\end{subequations}
With our assumptions, it is the Hilbert-Schmidt kernel, which ensures the uniqueness of the solution of Eq.~\eqref{Fredholm_eq}.

In what follows, we are interested not in the actual field $h(\mathbf{r})$, which is, due to the problem statement, a random function of the coordinates, but only in its statistical moments. In the general case, we assume that function $\zeta(x)$ is complex-valued and set its average value equal to zero, $\Av{{\zeta}(x)}=0$, while its binary correlation functions will be taken in the form
\begin{subequations}\label{Basic_corrs(x)}
\begin{align}
\label{zeta_zeta-corr(x)}
 & \Av{{\zeta}(x){\zeta}(x')} = \bm{\Xi}^2 W(x-x')\ ,\\
\label{zeta_zeta*-corr(x)}
 & \Av{{\zeta}(x)^*{\zeta}(x')} = |\bm{\Xi}|^2 W(x-x')\ .
\end{align}
\end{subequations}
The angle brackets $\Av{...}$ here and below denote statistical averaging over the realizations of random function $\zeta(x)$. Complex parameter $\bm{\Xi}=\sigma_R+i\sigma_I$ appearing in Eqs.~\eqref{Basic_corrs(x)} is called the (complex) variance of this function. The real parameters  $\sigma_{R,I}$ denote the variances of real and imaginary parts of the impedance, respectively. Function $W(x)$ is assumed to be real, even, normalized to unity at the maximum located at $x=0$, and sufficiently rapidly decaying to negligible values at the interval $|\Delta x|\sim r_c$ (correlation radius).

Equation \eqref{Fredholm_eq} can be formally solved with the use of the operator approach. With Fredholm operator $\hat{\mathcal{L}}$, the kernel of which is represented by Eq.~\eqref{Kernel_Fredholm_eq}, the solution of Eq.~\eqref{Fredholm_eq} can be written as
\begin{equation}\label{R(B)R(SPP)-oper_form}
  \widetilde{\mathcal{R}}(q)=
  -\mathrm{e}^{ik_{\text{\tiny SPP}}L/2}\bra{q}\big(\hat{\mathbf{1}}+
  \hat{\mathcal{L}}\big)^{-1}
  \left[t_-\hat{\mathcal{L}}\ket{k_{\text{\tiny SPP}}}+r_+\hat{\mathcal{L}}
  \ket{\!-\!k_{\text{\tiny SPP}}}\right]\ .
\end{equation}
Here we use Dirac notation for the matrix elements of the operator through bra- and ket-vectors. Operator $\hat{\mathcal{L}}$ appearing in Eq.~\eqref{R(B)R(SPP)-oper_form}), in its turn, is represented as a product of two operators, ${\hat{\mathcal{L}}=\hat{G}^{\scriptscriptstyle{(\text{CP})}}\hat{\zeta}_L}$, where the operator factors $\hat{G}^{\scriptscriptstyle{(\text{CP})}}$ and $\hat{\zeta}_L$ are given by matrix elements
\begin{subequations}\label{Oper_notations}
 \begin{align}
  \label{hatG(S)}
   & \bra{q}\hat{G}^{\scriptscriptstyle{(\text{CP})}}\ket{q'} =
   \left[\zeta_0+\sqrt{1-(q/k)^2}\right]^{-1}
   2\pi\delta(q-q')\ ,\\
  \label{hat_tilde_zeta}
   & \bra{q}\hat{\zeta}_L\ket{q'} =\widetilde{\zeta}(q-q')\ .
 \end{align}
\end{subequations}
The first of the above operators, $\hat{G}^{\scriptscriptstyle{(\text{CP})}}$, is the propagator of a certain wave excitation which will be denoted as a composite plasmon in what follows (the reader can found justification of this term in Ref.~\cite{bib:TarUsIak16}). The second factor, $\hat{\zeta}_L$, will be called the operator of the impedance perturbation.

Scattering amplitude \eqref{R(B)R(SPP)-oper_form} is written as a linear combination of matrix elements of a certain complex-valued operator
\begin{equation}\label{T=(1+L)^[-1]L}
  {\hat{\mathcal{T}}=\big(\hat{\mathbf{1}}+\hat{\mathcal{L}}\big)^{-1}\hat{\mathcal{L}}}\ .
\end{equation}
These matrix elements play the role of the propagators of the initial states described by vectors $\ket{\!\pm k_{\text{\tiny SPP}}}$ into the final state of scattering, which corresponds to vector $\ket{q}$. In this notation, the operator $\hat{\mathcal{L}}$ introduced in Eq.~\eqref{Kernel_Fredholm_eq} can be interpreted as the operator of one-fold scattering of the mode with an arbitrary $x$-component of the momentum from its ``initial'' state corresponding to the right mode index into the ``final'' (left) mode determined by propagator~\eqref{hatG(S)}. It is natural to associate the intensity of the scattering with the operator $\hat{\mathcal{L}}$ norm. Depending on the value of this norm, we will  define and consider the limiting cases of weak and strong intermode scattering below.

We define the norm of the operator $\hat{\mathcal{L}}$ in a standard way, in terms of the vector norm defined in corresponding Banach space,
\begin{align}\label{|A|^2}
  \|\hat{\mathcal{L}}\varphi\|^2=&\big(\hat{\mathcal{L}}\varphi,\hat{\mathcal{L}}\varphi\big)=
  \iint\limits_{-\infty}^{\quad\infty}\frac{d\kappa_1d\kappa_2}{(2\pi)^2}
  \widetilde{\varphi}^*(\kappa_1)\widetilde{\varphi}(\kappa_2)\nonumber\\
  &\times\int\limits_{-\infty}^{\infty}dx\iint_{\mathbb{L}}dx_1dx_2
  \bra{x}\hat{\mathcal{L}}\ket{x_1}^*\bra{x}\hat{\mathcal{L}}\ket{x_2}
  \mathrm{e}^{-i\kappa_1x_1+i\kappa_2x_2}\ .
\end{align}
Here $\varphi$ stands for a normalized trial function from the respective functional space. We assume that characteristic ``microscopic'' scales of spatial variation of this function are small as compared with length $L$, or at least do not exceed this length by order of magnitude. Given this assumption, it follows immediately that Fourier transforms of the trial functions alter significantly at characteristic scales large (or, at least, of the same order) in comparison with $1/L$, which is the characteristic width of prelimit $\delta$-functions arising when correlation relations \eqref{Basic_corrs(x)} are translated to momentum representation. Then, equality \eqref{|A|^2} is reduced to the following approximate form,
\begin{align}
\label{<|L|^2>-appr-2}
   \Av{\|\hat{\mathcal{L}}\varphi\|^2} &\approx |\bm{\Xi}|^2
   \int\limits_{-\infty}^{\infty}\frac{d\kappa}{2\pi}\left|\zeta_0+
   \sqrt{1-(\kappa/k)^2}\right|^{-2}
   \int\limits_{-\infty}^{\infty}\frac{d\kappa_1}{2\pi}\big|
   \widetilde{\varphi}(\kappa_1)\big|^2
   \widetilde{W}(\kappa-\kappa_1)\ .
\end{align}

The factors in the integral over $\kappa_1$ in Eq.~\eqref{<|L|^2>-appr-2}, both $\big|\widetilde{\varphi}(\kappa_1)\big|^2$ and $\widetilde{W}({\kappa-\kappa_1})$, have some scales of variation (localization). For function $\widetilde{\varphi}(\kappa)$ these may be $1/L$, $1/L^{\scriptscriptstyle{(\text{SPP})}}_{dis}$ and
$1/L^{\text{(loc)}}$, where under  $L^{\text{(loc)}}$ we mean the length of \emph{one-dimensional} Anderson localization, which coincides, by the order of magnitude, with the extinction length. All the above scales are small compared with the inverse correlation length $1/r_c$. This suggests that one can calculate the integral over $\kappa_1$ in Eq.~\eqref{<|L|^2>-appr-2} by taking correlation function $\widetilde{W}$ out of this integral at point $\kappa_1=0$. We thus arrive at equality
\begin{align}\label{L_norm-def-1}
  \Av{\|\hat{\mathcal{L}}\|^2}= |\bm{\Xi}|^2
  \int\limits_{-\infty}^{\infty}\frac{d\kappa}{2\pi}\left|\zeta_0+
  \sqrt{1-(\kappa/k)^2}\right|^{-2}
  \widetilde{W}(\kappa)\ ,
\end{align}
where function $\widetilde{W}(\kappa)$, based on the above made assumptions, may be taken out of the integral over $\kappa$ at the maximum points of the first integrand factor, viz., at $\kappa_{\pm}=\pm k'_{\text{\tiny SPP}}$. As a result, we obtain the following expression for the operator $\hat{\mathcal{L}}$ norm squared,
\begin{equation}\label{L_norm-def-2}
  \Av{\|\hat{\mathcal{L}}\|^2}\approx |\bm{\Xi}|^2
  \frac{k|\zeta_0''|}{\zeta'_0\sqrt{1+{\zeta_0''}^2}}
  \cdot\frac{1}{2}\sum_{\pm}\widetilde{W}(\pm k'_{\text{\tiny SPP}})\ .
\end{equation}
%

\section{Weak inter-mode scattering}
\label{Asymptota}

In the parameter range where the inequality holds
\begin{equation}\label{Weak_intermixing}
  \Av{\|\hat{\mathcal{L}}\|^2}\ll 1\ ,
\end{equation}
the scattering will be considered as weak. Expanding the inverse operator on the right-hand side of Eq.~\eqref{R(B)R(SPP)-oper_form} into the operator power series and retaining only the first term of this series, we can write the scattering amplitude in the following approximate form
\begin{align}\label{R(B)R(S)-|L|<<1}
  \widetilde{\mathcal{R}}(q)\approx -\mathrm{e}^{ik_{\text{\tiny SPP}}L/2}
  \bra{q}
  \left(t_-\hat{\mathcal{L}}\ket{k_{\text{\tiny SPP}}}+r_+\hat{\mathcal{L}}
  \ket{\!-\!k_{\text{\tiny SPP}}}\right)\ .
\end{align}
Obviously, in Eq.~\eqref{R(B)R(S)-|L|<<1} only one-fold scattering through composite plasmons is taken into account.

\begin{figure}[h!!]
  \centering
  \scalebox{.4}[.4]{\includegraphics{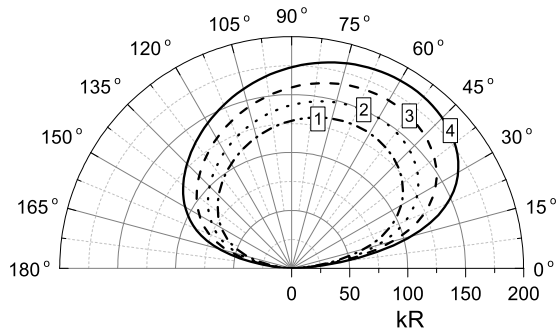}}
  \caption{Radiation patterns for $\|\hat{\mathcal{L}}\|\ll 1$ and different values of the real part of the surface impedance. Curve 1 --- $\zeta'_0=0,1$; curve 2 --- $\zeta'_0=0,07$; curve 3 --- $\zeta'_0=0,04$; curve 4 --- $\zeta'_0=0,01$. Other parameters are identical for all the diagrams: $\zeta''_0=-0,25$; $\Xi=0,125$; $kr_c=0,5$; $kL=7$.
  \hfill\label{fig2}}
\end{figure}

The result obtained thereby consists in that the parent plasmon-polariton incident onto the surface region with modulated impedance passes through this region being scattered into the bulk modes relatively weakly. In this work, we do not dwell on this limiting case, referring the interested reader to paper~\cite{bib:TarUsIak16}. We only provide here in figure~\ref{fig2} the radiation diagram obtained in that paper.
Note especially that if the dissipation in the metallic part of the considered medium is gradually reduced, the PPW scattering into the upper (dielectric) half-space in the form of spatial harmonics quasi-isotropically distributed over the angle variable increases, while it does not undergo essential variation within the plane of the primordial PPW propagation.

\section{Strong mixing of scattered modes}
\label{Asymptota2}

The scattering cannot be considered as weak, and, correspondingly,
one cannot limit oneself to single scattering, if the norm of the
operator $\hat{\mathcal{L}}$ becomes of the order of unity, or a
fortiori when it exceeds it. We will consider the case where the
inequality holds
\begin{equation}\label{Strong_intermixing}
  \Av{\|\hat{\mathcal{L}}\|^2}\gg 1
\end{equation}
as the limit of strong inter-mode scattering. If the above inequality is satisfied, expression~\eqref{R(B)R(SPP)-oper_form} for the scattering amplitude can be rearranged by simple algebraic manipulation with operator~\eqref{T=(1+L)^[-1]L} and converted to the following approximate (in parameter~$\|\hat{\mathcal{L}}\|^{-1}$) form,
\begin{align}
\label{R(B)R(SPP)-inverse_L}
  \widetilde{\mathcal{R}}(q) & =-\mathrm{e}^{ik_{\text{\tiny SPP}}L/2} \bra{q}\big(\hat{\mathbf{1}}+\hat{\mathcal{L}}\big)^{-1}\left[t_-\hat{\mathcal{L}}\ket{k_{\text{\tiny SPP}}}+r_+\hat{\mathcal{L}}\ket{\!-k_{\text{\tiny SPP}}}\right]
\notag\\
  &=-\mathrm{e}^{ik_{\text{\tiny SPP}}L/2}
  \left[\bra{q}\hat{1}-\big(\hat{\mathbf{1}}+\hat{\mathcal{L}}\big)^{-1}\right]
  \Big[t_-\ket{k_{\text{\tiny SPP}}}+r_+\ket{\!-k_{\text{\tiny SPP}}}\Big]
\notag\\
 & \approx -\mathrm{e}^{ik_{\text{\tiny SPP}}L/2}\Big\{t_-\Delta_L(q-k_{\text{\tiny SPP}})
  +r_+\Delta_L(q+k_{\text{\tiny SPP}})\notag\\
&\phantom{\approx\ } -\bra{q}\hat{\mathcal{L}}^{-1}\Big[t_-\ket{k_{\text{\tiny SPP}}}+r_+\ket{\!-k_{\text{\tiny SPP}}}\Big]\Big\}\ .
\end{align}

From Eq.~\eqref{R(B)R(SPP)-inverse_L}, in the leading approximation in $\|\hat{\mathcal{L}}\|^{-1}\ll 1$ the scattered field is
\begin{align}\label{BS_fields-strong_scatt}
  h^{(0)}(\mathbf{r}) =-\mathrm{e}^{ik_{\text{\tiny SPP}}L/2}
  \int\limits_{-\infty}^{\infty}\frac{dq}{2\pi}
  \Big[t_-\Delta_L(q- & k_{\text{\tiny SPP}})+r_+\Delta_L(q+k_{\text{\tiny SPP}})\Big]\nonumber\\
  &\times\exp\left(iqx+i\sqrt{k^2-q^2}z\right)\ .
\end{align}
Setting $z=0$ in Eq.~\eqref{BS_fields-strong_scatt}, we immediately obtain the value of the scattered field at the metal-insulator interface inside segment $\mathbb{L}$,
\begin{equation}\label{h(0)_strsc-z=0}
   h^{(0)}(x,0) \approx -\mathrm{e}^{ik_{\text{\tiny SPP}}L/2}
   \left(t_-\mathrm{e}^{ik_{\text{\tiny SPP}}x}+
   r_+\mathrm{e}^{-ik_{\text{\tiny SPP}}x}\right)
   \qquad(x\in\mathbb{L})\ .
\end{equation}
Matching the values of this field with the zero values outside segment $\mathbb{L}$, we obtain a set of two equations for determining the coefficients $t_-$ and $r_+$. If the lower medium is dissipative and the wave number $k_{\text{\tiny SPP}}$ is, accordingly, complex, the system has only the trivial solution, $t_-=r_+=0$. Instead, if we assume zero dissipation, a nontrivial solution with arbitrary $t_-=\pm r_+$ becomes possible. In this case the polariton propagation parameter is quantized through the condition $k_{\text{\tiny SPP}}L=n\pi$, which corresponds to the formation of a~resonator (standing waves) inside disordered segment of the interface. Formally, the resonant reflection occurs due to the mismatch of the wave states within the ``disordered'' segment and outside it~\cite{RemarkTarShost15}. But regardless of whether or not there is dissipation in the conductor, in both cases the matching conditions \eqref{Joining_Eqs} lead to the equations
\begin{equation}\label{Entire_reflection}
  1+r_- =0\ ,  \quad t_+ =0\ ,
\end{equation}
meaning that in the leading approximation in the inverse norm of the
operator~$\mathbb{L}$, the segment of the boundary with the
perturbed impedance acts as an almost perfect reflector for the
incident PPW.

In the region $kz\gg 1$, where it does not altogether make sense to
talk about plasmon-polaritons, the scattered field can be calculated
with asymptotic accuracy far away from the scattering source,
i.\,e., the segment of the perturbed boundary. Assuming that
inequality $|\mathbf{r}|=\sqrt{x^2+z^2}\gg L$ is fulfilled and using
the stationary phase method we arrive at the following formula for
the scattered field in the Fraunhofer zone,
\begin{align}\label{h(0)(kz>>1)}
  h^{(0)}(\mathbf{r})\approx -\sin\phi\sqrt{\frac{2\pi k}{R}}
  &  \exp\Big[i\left(kR+k_{\text{\tiny SPP}}L/2-\pi/4\right)\Big]\nonumber\\
  &\times\Big[t_-\Delta_L(q_0-k_{\text{\tiny SPP}})+r_+\Delta_L(q_0+k_{\text{\tiny SPP}})\Big]\ .
\end{align}
Here, $q_0=kx/R=k\cos\phi$ is the stationary phase point. Since accounting for even weak dissipation in the conductor leads to the vanishing coefficients $t_-$ and $r_+$, the field \eqref{h(0)(kz>>1)}, which forms the scattering diagram, is zero in the leading approximation. The suppression of the leaking waves, in conjunction with Eqs.~\eqref{Entire_reflection}, leads to an effectively one-dimensional scattering of surface polaritons in an open and essentially non-1D system. Yet, one should keep in mind that this effect can only occur under strong scattering of the PPW, which actually corresponds to inequality
\begin{equation}\label{StrongScatt-simpl}
  |\bm{\Xi}|^2k^2L^{\scriptscriptstyle{(\text{SPP})}}_{dis}\widetilde{W}(k'_{\text{\tiny SPP}})\gg 1\ .
\end{equation}
This relation is certainly satisfied in the dissipation-free limit,
since in this case
${L^{\scriptscriptstyle{(\text{SPP})}}_{dis}\to\infty}$.

\section{Discussion}

To conclude, in this study a detailed theory of the scattering of plasmon-polariton waves by a segment of the metal-dielectric (vacuum) interface with randomly distributed surface impedance is developed. The applied operator approach made it possible to analyze the scattered fields for any scattering intensity, from weak scattering, commonly considered in the other theories, to extremely strong scattering. To estimate the level of scattering we proposed and justified a suitable criterion based on the Hilbert norm of the mixing operator for the intermediate scattering states --- composite polaritons. The term composite refers to the mixed states, which represent a weighted sum of true surface polaritons and specific bulk (quasi-Norton) waves responsible for the energy leakage into the dielectric upon scattering of the PPW by the defective segment of the interface.

Under conditions when the plasmon-polariton scattering is regarded as weak, which is only possible in the case of a sufficiently high level of dissipation in the conductor, the main result of the scattering, except for the dissipative loss, is the radiation of some fraction of PPW energy from the surface of the conductor into the bulk of the dielectric. We have succeeded in calculating the radiation pattern and shown that the radiated energy is proportional to the Fourier transform of the binary correlation function of the impedance, which is taken at a point equal to the difference between the wave number of the surface polariton and the projection on the boundary plane of the wave vector
of quasi-Norton harmonic directed towards the (remote) position of the detector (see Ref.~\cite{bib:TarUsIak16} for details). This dependence has an obvious resemblance to the angular dependence of the fields scattered by reflective gratings, in which Wood~\cite{Wood1902} once discovered the anomalies in the form of reflection resonances and the origin of which has been later explained by Fano~\cite{Fano41}. The role of such resonances in the PPW scattering by an interface segment with random impedance should be played by the maxima of the impedance correlation function. For generic random processes such maxima are usually located near zero argument of the correlation function, and the above-mentioned difference between the wavenumbers typically does not coincide with them.

If the parameters of the impedance and the plasmon-polariton wave are such that the scattering is not weak but rather is strong (in the sense of inequality~\eqref{StrongScatt-simpl}), the result of the scattering is reduced to almost mirror reflection of the incident surface polariton and the suppression of the quasi-Norton (bulk) component of its radiation. It should be noted that the PPW scattering can become strong not only through an increase in the impedance variance, but also due to the reduction in the dissipative loss in the underlying conductor. The norm of the mixing operator for the scattered modes is inversely proportional to the dissipative part of the impedance and can become arbitrarily large even for small but finite values of its reactive component. The situation here is reminiscent of that occurring at normal incidence of a plane wave on a semi-bounded one-dimensional disordered medium, in which, due to the Anderson (interference) localization of states, further penetration of the field becomes impossible. The incident wave is completely (yet again, in the absence of dissipative losses) reflected in backward direction.

Meanwhile, the propagation medium we have chosen for analysis is not one-dimensional. Moreover, the system considered here is non-Hermitian, and in such systems the Anderson localization is known to be, as a rule, rather weak. The effective ``one-dimensionalization'' of the system and, as a consequence, the absence of waves leaking to the upper half-space arises in our case as a~result of wave interference upon their \emph{multiple} scattering on the impedance fluctuations. In view of this, the fact that the effectiveness of the interference, along with the intensity of the scattering, reduces with the increase in the dissipation within the conducting substrate seems to be completely natural.

The fact that under strong scattering the plasmon-polariton energy leakage into the bulk of the dielectric is suppressed should, of course, play a positive role in plasmonics. Purposeful turning the obstacles along the propagation path of the surface wave into reflectors allows to effectively control its propagation direction, hence creating open-type surface waveguides.




\begin{thebibliography}{cc}
\bibitem{RemarkTarShost15}
Resonance oscillations of a similar nature have been described in Ref.~\cite{TarShost15}, where the problem of wave propagation in waveguides containing a segment with weakly corrugated boundaries was solved.
\end{thebibliography}
\end{document}